\documentclass[a4paper,10pt,leqno]{article}

\usepackage{mathtools}
\usepackage{natbib}

\begin{document}
	\setcounter{page}{1}
	
	\title{A Note on the Moments of Special Mixture Distributions, with Applications for Control Charts}
	
	\author{{\bf Bal\'{a}zs Dobi, Andr\'{a}s Zempl\'{e}ni} \\Budapest, Hungary\\[1ex]}
	%% More authors can be given by repeting the \author comand)

	%

	%

	%
	
	%\projsupport{The work was supported by the project EFOP-3.6.2-16-2017-00015, which was
	%	supported by the European Union and co-financed by the European Social Fund.}
	%
	
	%\begin{center}
	%\textit{Dedicated to ...}
	%\end{center}

	%%%%%%%
	% We shall fill out "???"
	%%%%%%%%

%	\vspace{-7ex}
	\maketitle

	Abstract: {Control charts can be applied in a wide range of areas,
		this paper focuses on generalisations suitable for healthcare applications. We concentrate on the effect of using mixture distributions as the possible shifts in the process mean value. We give a closed formula for the  expected squared deviation from the target value between samplings. This observation can be utilised in finding the cost-optimal parameters of the control chart.}
	\\

	
	\section{Introduction}
	
	Control charts, as methods of statistical process control in industry have been introduced by Walter A. Shewhart in the 1920s \citep{Montgomery:2009}. Even though initially control charts were optimised with respect to statistical criteria, the concept of cost-efficient or cost-optimal control charts appeared not long after. One of the earliest and most important work in this field is A. J. Duncan's paper from 1956 \citep{duncan1956economic}. The concept is still very popular today as can be seen by published articles and developed software packages (see e.g. \cite{mortarino2010duncan,Zhu+Park:2013}).
	
	 Our earlier research focused on applications in healthcare settings, see e.g. \cite{Dobi+Zempleni:2019a}.	
	In this paper, the main model is a flexible framework which is based on the works of \cite{Zempleni+etal:2004}. This framework uses a Markov chain-based approach, which is similar to Duncan's cycle model as it also defines states related to the monitored process. The advantage of this approach was that it allowed generalisations for random shift sizes, random repairs and random sampling times, all of which are common in healthcare applications. Using these control charts, we were able to estimate the optimal parameters of a patient monitoring setup, which consisted of the optimal time between samplings (i.e. control visits) and critical value (i.e. medical criteria) \citep{Dobi+Zempleni:2019a}.
	
	In this paper we aim to assess the effect of mixture shift size distributions - distributions which model the degradation in quality - on the average cost over a given time period. We give a closed formula for one of the main terms in the cost-calculations for control charts.
	
	The rest of the paper consists of the following parts: Section 2 contains the main results, but first the mathematical background for the later results are given. Namely Subsection 2.1 briefly discusses the Markov chain-based framework and its generalisations. Subsections 2.2 and 2.3 deal with the main subject of this note: we present here a formula for the squared expectation of mixture shift distributions over a given time interval. Section 3 concludes the paper.

	%----------------------------------------------------------------------------------------
	%	Methods
	%----------------------------------------------------------------------------------------
	
	\section{Methods}
	The description of the Markov chain-based framework below is just a brief introduction and summary necessary for understanding the results presented later in the paper. For further reading and more detailed descriptions see \cite{Zempleni+etal:2004,Dobi+Zempleni:2019a,Dobi+Zempleni:2019b,Dobi+Zempleni:2022}.
	
	%------------------------------------------------
	\subsection{The Markov chain-based framework and its generalisations}
	Our control charts are used to monitor a 
	characteristic, based on a single sample element at a time. The shift in this characteristic - when it occurs - is positive. These assumptions correspond to an $X$-chart setup with sample size $N=1$ and one-sided critical value $K$. Usually, there are three free parameters which are the focus of optimisation: the sample size $N$, the critical value $K$, and the time between samplings $h$. Since the sample size is fixed here as $N=1$, we are left with two parameters to use: $K$ and $h$.
	
	The following parameters and constants used throughout the paper are supposed to be known:
	
	\begin{center}
		\begin{tabular}{||l l||} 
\hline
\hline
			\textbf{Notation} & \textbf{Meaning}  \\ [1ex] 
\hline
\hline
			$\mu_0$ & Target value, in-control expectation
			 \\ [1ex] 
			\hline
			$\sigma$ & Process standard deviation
			 \\ [1ex] 
			\hline
			$\delta$ & Expected value of the exponential distribution (shift size)  \\ [1ex] 
			\hline
			$\zeta$ & Probability of geometric shift in the mixed distribution  \\ [1ex] 
			\hline
			$\xi$ & Probability parameter of the geometric distribution  \\ [1ex] 
			\hline
			$s$ & Expected number of shifts in a unit time \\ [1ex] 
				\hline
			$c_s$ & Sampling cost \\ [1ex] 
			\hline
			$c_{os}$ & Base out-of-control cost per unit time
			 \\ [1ex] 
			\hline
			$c_{ob}$ & Shift-proportional out-of-control cost per unit time
			\\ [1ex] 
			\hline
			$c_{rb}$ & Base repair cost \\ [1ex] 
			\hline
			$c_{rs}$ & Shift-proportional repair cost \\ [1ex] 
			\hline
			\hline
		\end{tabular}
	\end{center}
	The followings are also assumed:
	\begin{itemize}
		\item The measurement error is normal with expectation 0 and known standard deviation $\sigma$. Let its cumulative distribution function (CDF) be denoted by $\phi$. Thus the in-control process distribution is normal with parameters $\mu_0$ and $\sigma$.
		\item The shift intensity ($1/s$: the inverse of the expected number of shifts in a unit time) is constant. 
		\item The shift size distribution is assumed to be known. 
		\item The process does not repair itself, and when a repair (treatment) is carried out it does not repair the process perfectly. Furthermore, the repair itself is treated as
		an instantaneous event, thus all costs related to repairing should be included in the repair cost. For example if a major repair entails higher cost, then this should also be	reflected in the calculation. 
		\item The time between shifts, the shift sizes and repair effectiveness are all assumed to be independent from each other.
	\end{itemize}
	Using the above assumptions, the future distances from $\mu_0$ are only
	dependent on the current distance and independent from the earlier ones. As we consider the states at the samplings, the process is a Markov chain, with a continuous state space. In our earlier papers we introduced a discretisation scheme, which was needed in the implementation of the algorithm. In this paper we do not consider this in detail, as the theoretical results will not need discretisation, only the general form of the cost function.
	
	%------------------------------------------------
	Our approach contains two important differences compared to the traditional models. First is the random shift size, introduced for the cost-optimal approach by \cite{Zempleni+etal:2004}. The second is the random repair size, introduced by \cite{Dobi+Zempleni:2019a}. In this note we focus on the first one.
	
	Let $\tau_i$ denote the random shift times on the
	real line and let $\rho_i$ be the shift size at time $\tau_i$.  Assume that $\rho_i$ follows a  distribution, which
	has a CDF with support over $(0, \infty)$, and that
	the shift sizes have the same distribution, are independent from each other and from $\tau_i$. Let the probability mass function (PMF) of the number of shifts after time $t$ from the 
	process' start (assumed to be in the in-control state)
	be denoted by $\nu_t$. $\nu_t$ is a discrete distribution with support over
	$\mathbf{N}^0$, which will be assumed to have the Poisson distribution with parameter $\lambda t$.
	If the previous conditions are met, the resulting random process of the shifts - let us denote it by $H(t)$ - has step functions as trajectories, which are monotonically increasing between samplings. The
	CDF of the process values for a given time $t$
	from the start can be written the following way (assuming that there was no alarm signal and repair before $t$):
	\begin{equation} \label{shiftd}
	Z_t(x)=
	\begin{cases}
	0 &{\text{if }} x < 0, \\
	\nu_t(0) + \sum_{k=1}^\infty \nu_t(k) \Psi_k(x) &{\text{if }} x \geq 0,
	\end{cases}
	\end{equation}
	where $\Psi_k()$ is the CDF of the
	sum of $k$ independent, identically distributed shift sizes $\rho_i$.
	The case $x=0$ means there is no shift, and the probability of zero shift size is
	just the probability of no shift occurring, which is $\nu_t(0)$.

	The resulting process of 
	the expected values is a monotone increasing step function between samplings and has a
	downward "jump" at alarm states - as the repair is assumed to be
	instantaneous.
	
	The form of the cost function we used is
	\begin{equation} \label{EC}
		\mathrm{E}(C)=C \cdot P,
	\end{equation}
	where $E(C)$ is the expected cost per unit time, $C$ is a vector of the total cost per unit time associated with each state and $P$ is the stationary distribution (vector of probabilities), defined by a suitable discretisation scheme. $E(C)$ is the expected value of the discrete distribution where the possible values are the costs related to the states and their probabilities are set by the stationary distribution of the Markov chain.
	
	The $C$ vector is
	$$C=\Bigg\{\frac{c_s}{h},\frac{c_s+c_f}{h},\frac{c_s}{h}+c_o,\frac{c_r+c_s}{h}+c_o B_h\Bigg\},$$
where $c_s$ is the sampling cost, $c_f$ is the false alarm cost, $c_r$ is the true alarm (repair) cost and $c_o$ is the out-of-control operation cost per unit time. This cost item will be investigated in more detail in the sequel.

The relationship between the distance from $\mu_0$ and the resulting costs are often based on a Taguchi-type loss function \citep{Montgomery:2009}. This means that a squared loss function is assumed. 
	The expected out-of-control operation cost can be written as the expectation of
	a function of the distance from the target value. At (\ref{shiftd}) the shift size
	distribution was defined for a given time $t$, but this time we are
	interested in calculating the expected cost for a whole interval. Formally, we intend to calculate the area ${\cal C}_{h,j}^2$ under the curve $\{t,E((H_{t_0,j}(t))^2)\}=\int_{t_0}^{t_0+h}E((H_{t_0,j}(t))^2)dt$, where $t_0\le t\le t_0+h$ and $H_{t_0,j}$ is the shifted mean-process (its unconditional distribution function $Z_t$ was given in (\ref{shiftd})) upon the condition that $H(t_0)=j$. Note that the area under the curve is the total cost generated in a time interval, the expected cost per unit time for practical use can be calculated from this value by simply dividing with the length of the interval.
	
  Thus, the main question here is the integral of the expected squared shift size over a given interval. This is the subject of the next subsection, in the case of a mixture distribution.
	
\subsection{Mixture distribution as the shift size}

For practical demonstration purposes, let us assume first that the shift times form a homogeneous Poisson process, and the size of a single shift has a distribution which is a mixture of an exponential and a geometric distribution:
$$F_m(x)=\zeta F_g(x)+(1-\zeta )F_e(x),$$
where $\zeta \in [0,1]$ is the mixing parameter, $F_g()$ is the CDF of the geometric distribution, $F_e()$ is the CDF of the exponential distribution, and $F_m()$ denotes the CDF of the resulting mixture distribution. Such a distribution can model processes with slow degradation, mixed with sudden jumps. A real-life example could be the effectiveness of a professional athlete, where the continuous degradation can be attributed to age and sudden degradation to accidents. Another example could be simultaneously occurring chronic kidney disease and acute kidney injury \citep{Hsu2016role}. Different definitions are available for the geometric distribution, the one used here has support over positive integers, thus $F_g(x)=1-(1-\xi)^{x}$, with probability parameter $\xi$. A negative binomial distribution can be defined as the convolution of independent geometric distributions with the same parameter. Its PMF can be written in the following way:
\begin{equation} \label{negbindef}
f_{nb}(x)={x-1 \choose x-r}\cdot \xi^{r}(1-\xi)^{x-r},
\end{equation}
where $r$ is the number of summed geometrically distributed variables. The support of this distribution is $x \in \{r,r+1,r+2,...\}$. Let us assume now that $n$ shifts occurred in a given time interval, and $r$ of these were geometrically, while the rest were exponentially distributed. The CDF of the distribution of the sum of the shifts is then
\begin{equation}
F_{M|n,r}(x)=
\begin{dcases}
	F_{E_{n-r}}(x) &{\text{if }} r=0, \\
	\sum_{l=r}^{\left\lfloor x \right\rfloor} F_{E_{n-r}}(x-l)f_{nb}(l) &{\text{otherwise}},
\end{dcases} \nonumber
\end{equation}
where $F_{E_{n-r}}$ is the distribution function of the Erlang distribution
$E(n-r,\frac{1}{\delta})$. The distribution function of the sum of $n$ variables with exponential-geometric mixed distribution is then
\begin{equation}
F_{M|n}(x)=\sum_{r=0}^n f_b(r)F_{M|n,r}(x), \nonumber
\end{equation}
where $f_b()$ is the PMF of the binomial distribution with parameters $n$ and $\zeta$, giving the probability of $r$ geometrically distributed shifts out of $n$.

If one wants to construct the shift size distribution for a given time after start using the notations in (\ref{shiftd}), $\nu_t$ would be the PMF of the Poisson distribution, and $\Psi_k()$ would be $F_{M|n}()$ with $n=k$.

For the following paragraphs, let $X$ denote an $E(u,\frac{1}{\delta})$ (Erlang) distributed random variable, which is the sum of $u$ independent exponential variates each with mean $\delta$, and let $Y$ denote a $NegBin(u,\xi)$ (negative binomial) distributed random variable, which is the sum of $u$ independent geometric variates each with parameter $\xi$. Note, that for the negative binomial distribution we are using the definition given at (\ref{negbindef}). Also, for practical use, one may wish to have discrete jump size other than the integers defined by the support of the negative binomial distribution. We can simply multiply the random variable by a constant (let us denote it by $J$) to create arbitrary jump-intervals. The initial equation thus takes the form
\begin{equation}
{\cal C}_{h,j}^2 = \int_{0}^{h}  \Bigg[e^{-ts}j^2 +
\Bigg(\sum_{k=1}^\infty \frac{(ts)^k e^{-ts}}{k!} \cdot
E((X+JY+j)^2)
\Bigg) \Bigg] dt\nonumber \\
\end{equation}
In this equation $j$ represents an arbitrary starting distance greater than or equal to $\mu_0$, since the process may not start at $\mu_0$ after a sampling. The innermost expectation, which was in \cite{Dobi+Zempleni:2019a} simply the expectation of $(X+j)^2$, now is replaced by $E((X+JY+j)^2)$. It is easier to present the calculation in parts, as it will result in a long expression. Firstly,
\begin{align}
E((X+JY+j)^2) = \sum_{r=0}^{k} \binom{k}{r} \zeta^r (1-\zeta)^{k-r} &\Bigg[\Bigg(\frac{k-r}{1/ \delta}\Bigg)^2+\frac{k-r}{1/\delta^2}+ \\
& \ + \Bigg(J \frac{r}{\xi}\Bigg)^2+J^2 r \frac{1-\xi}{\xi^2}+j^2 + \nonumber \\
& \ + 2 \frac{k-r}{1/\delta} J \frac{r}{\xi}+ 2 \frac{k-r}{1/\delta} j + 2 J \frac{r}{\xi} j\Bigg]. \nonumber
\end{align}
Notice that we only expanded $E(X+JY+j)^2$ based on the well-known form of $(a+b+c)^2$, operating with the constant $j$ as a random variable with degenerate distribution, and using the independence between variables. Namely, for fixed $k$ and $r$, $\Big(\frac{k-r}{1/ \delta}\Big)^2+\frac{k-r}{1/\delta^2}$ is $E(X^2)$, $\Big(J \frac{r}{\xi}\Big)^2+J^2 r \frac{1-\xi}{\xi^2}$ is $E(Y^2)$. The rest of the terms are just $2E(XY)$, $2E(Xj)$ and $2E(Yj)$. The number of geometric variables follow a binomial distribution, hence the formula in the beginning of the right hand side of the equation. $k$ in this equation is a positive integer constant, but otherwise is modelled by a Poisson distribution. The expectation is reduced to
\begin{align}E((X+JY+j)^2) & =  k\delta^2 + (j - k\delta(\zeta - 1))^2 + \\
& \ + \frac{2 k \zeta J(j+\delta (k+\zeta-k \zeta-1)) - \xi k(\delta \zeta)^2}{\xi}+ \nonumber \\
& \ + \frac{k \zeta J^2(2-\xi+\zeta(k-1))}{\xi^2} \nonumber
\end{align}
Notice, that the result cannot be simply partitioned into a purely Erlang or a purely negative binomial part. The rest of the calculation is the same as for the exponential (non-mixture) shift size case in \cite{Dobi+Zempleni:2019a}, albeit with more complicated expressions:
\begin{align}
{\cal C}_{h,j}^2 &= \int_{0}^{h}  e^{-ts}j^2 + \sum_{k=1}^\infty \frac{(ts)^k
	e^{-ts}}{k!}\big(E((X+JY+j)^2)\big) dt  \\
&= \int_{0}^{h} j^2+\frac{2 j s t (\delta (\xi-\xi \zeta)+\zeta J)}{\xi}+ \nonumber \\
& \ +\frac{s t (\delta^2 \xi^2 (\zeta-1) (-2+ s t (\zeta-1))-2 \delta \xi \zeta s t J (\zeta-1)+\zeta (2-\xi+\zeta s t) J^2)}{\xi^2}dt \nonumber \\
&=j^2+\delta h j s-\delta h j \zeta s+\frac{h j \zeta s J}{\xi}+ \nonumber \\
& \ +\frac{h s (-6 \delta^2 \xi^2 (\zeta-1)-3 (\xi-2) \zeta J^2+2 h s (\delta (\xi-\xi \zeta)+\zeta J)^2)}{6 \xi^2} \nonumber
\end{align}

The generality of the proposition is highlighted by its application on mixture distributions, since there is no readily-available density function to be used to ease the calculations, like in the simpler, exponential case, treated e.g. in \cite{Dobi+Zempleni:2019a}.

\subsection{General form for 2-component mixture shift size distributions}

The case above demonstrates that ${\cal C}_{h,j}^2$ can be calculated using a closed form for non-trivial shift size distributions too. Several assumptions were made for the special cases so far, however the results can be generalised. Let us now only assume that $X$ and $Y$ are random variables with finite expectation and variance. Note that $X$ and $Y$ are not the component variates themselves, but the resulting variables after summing the stacked shifts. Let us denote these expectation and variance values with $m_X$, $v_X$, $m_Y$ and $v_Y$ for $X$ and $Y$ respectively. The calculation of the general form is very similar to the specific case above:
\begin{equation}
{\cal C}_{h,0}^2 = \int_{0}^{h}
\Bigg(\sum_{k=1}^\infty \frac{(ts)^k e^{-ts}}{k!} \cdot
E((X+Y)^2)
\Bigg) dt\nonumber \\
\end{equation}
$E((X+Y)^2)$ takes on the following form:
\begin{align}
E((X+Y)^2) = \sum_{r=0}^{k} \binom{k}{r} \zeta^r (1-\zeta)^{k-r} &\Bigg[[(k-r)m_X]^2+(k-r)v_X \\
& \ +[r m_Y]^2+r v_Y + \nonumber \\
& \ + 2(k-r)m_X r m_Y\Bigg]. \nonumber
\end{align}

The rationale behind the formulation is simply that a variable which is defined as the sum of $(k-r)$ $X$ variates will have the expectation $(k-r)m_X$ and variance $(k-r)v_X$, since we assumed finite expectation and variance for $X$. The same is true for a variable that is the sum of $r$ $Y$ variates. The calculation above results in
\begin{align}
E((X+Y)^2) & = k \big[m_X^2 k + v_X - \zeta v_X + \\
& \ + \zeta \big(v_Y-(m_X-m_Y) (m_Y+m_X (2 k-1))\big) + \nonumber \\
& \ + \zeta^2 (m_X-m_Y)^2 (k-1)\big]. \nonumber
\end{align}
Again, the rest of the calculation is quite straightforward, as we only have to repeat the steps well-established above:
\begin{align}
{\cal C}_{h,0}^2 &= \int_{0}^{h}  \sum_{k=1}^\infty \frac{(ts)^k
	e^{-ts}}{k!}\big(E((X+Y)^2)\big) dt  \\
&= \int_{0}^{h} s t (m_X^2+v_X-m_X^2 \zeta-v_X \zeta+m_Y^2 \zeta+v_Y \zeta+(m_X-m_X \zeta+m_Y \zeta)^2 s t) dt \nonumber \\
&= \frac{1}{6} h^2 s (3 (m_X^2+v_X + \zeta (m_Y^2+v_Y-m_X^2-v_X))+2 h (m_X-\zeta(m_X-m_Y))^2 s).  \nonumber
\end{align}

From this general equation, one can easily get the previously-presented \citep{Dobi+Zempleni:2019a} exponential shift size case for $\zeta=0$, $j=0$, $m_X=\delta$, $v_X=\delta^2$:
\begin{align}
&\frac{1}{6} h^2 s (3 (m_X^2+v_X + \zeta (m_Y^2+v_Y-m_X^2-v_X))+2 h (m_X-\zeta(m_X-m_Y))^2 s) \\
&= \frac{1}{6} h^2 s (3 (\delta^2+\delta^2 + 0 (m_Y^2+v_Y-\delta^2-\delta^2))+2 h (\delta-0(\delta-m_Y))^2 s) \nonumber \\
&= \frac{1}{6} h^2 s (6 \delta^2 + 2 h \delta^2 s)= h^2s\delta \Big(\delta + \frac{hs\delta}{3}\Big). \nonumber
\end{align}

It is also useful to inspect the results, if $m_X=m_Y$ and $v_X=v_Y$:
\begin{align}
&\frac{1}{6} h^2 s (3 (m_X^2+v_X + \zeta (m_Y^2+v_Y-m_X^2-v_X))+2 h (m_X-\zeta(m_X-m_Y))^2 s) \\
&= \frac{1}{6} h^2 s (3 (m_X^2+v_X + \zeta (m_X^2+v_X-m_X^2-v_X))+2 h (m_X-\zeta(m_X-m_X))^2 s) \nonumber \\
&= \frac{1}{6} h^2 s \Big(3 (m_X^2+v_X) + 2 h s m_X^2\Big). \nonumber
\end{align}
One can see that $\zeta$ disappears from the equation, as expected.

	\section{Conclusions}
	In this note we presented a general formula for the expectation of the squared shift, when the shift size is a mixture distribution. This speeded up the\texttt{ Markov- chart R} package, as shown in \cite{Dobi:2020package,Dobi+Zempleni:2022} and has potential applications for other important cases, like normal and Frechet mixtures - where both components are continuous, but we face a mixture of a light tailed and a heavy tailed shift. We plan to come back to the implementation of these results later.

\bibliographystyle{plainnat}
\bibliography{refs}

\begin{thebibliography}{10}
\providecommand{\natexlab}[1]{#1}
\providecommand{\url}[1]{\texttt{#1}}
\expandafter\ifx\csname urlstyle\endcsname\relax
  \providecommand{\doi}[1]{doi: #1}\else
  \providecommand{\doi}{doi: \begingroup \urlstyle{rm}\Url}\fi

\bibitem[Dobi and Zempl\'{e}ni(2019{\natexlab{a}})]{Dobi+Zempleni:2019a}
Bal\'{a}zs Dobi and Andr\'{a}s Zempl\'{e}ni.
\newblock Markov chain-based cost-optimal control charts for health care data.
\newblock \emph{Quality and Reliability Engineering International}, 35\penalty0
  (5):\penalty0 1379--1395, 2019{\natexlab{a}}.
\newblock \doi{10.1002/qre.2518}.

\bibitem[Dobi and Zempl\'{e}ni(2019{\natexlab{b}})]{Dobi+Zempleni:2019b}
Bal\'{a}zs Dobi and Andr\'{a}s Zempl\'{e}ni.
\newblock Markov chain-based cost-optimal control charts with different shift
  size distributions.
\newblock \emph{Annales Universitatis Scientiarum Budapestinensis de Rolando
  E\"{o}tv\"{o}s Nominatae, Sectio Computatorica}, 49:\penalty0 129--146,
  2019{\natexlab{b}}.
\newblock URL
  \url{\texttt{http://ac.inf.elte.hu/Vol{\_}049{\_}2019/129{\_}49.pdf}}.

\bibitem[Dobi and Zempl{\a'e}ni(2020)]{Dobi:2020package}
Bal{\a'a}zs Dobi and Andr{\a'a}s Zempl{\a'e}ni.
\newblock \emph{\texttt{Markovchart}: Markov Chain-Based Cost-Optimal Control
  Charts}, 2020.
\newblock URL \url{https://CRAN.R-project.org/package=Markovchart}.
\newblock \texttt{R}~package version~2.1.4.

\bibitem[Dobi and Zempl\'{e}ni(2022)]{Dobi+Zempleni:2022}
Bal\'{a}zs Dobi and Andr\'{a}s Zempl\'{e}ni.
\newblock Markovchart: An {R} package for cost-optimal patient monitoring and
  treatment using control charts.
\newblock \emph{Computational Statistics}, 2022.
\newblock \doi{10.1007/s00180-021-01175-3}.

\bibitem[Duncan(1956)]{duncan1956economic}
Acheson~J Duncan.
\newblock The economic design of x charts used to maintain current control of a
  process.
\newblock \emph{Journal of the American Statistical Association}, 51\penalty0
  (274):\penalty0 228--242, 1956.
\newblock \doi{10.1080/01621459.1956.10501322}.

\bibitem[Hsu and Hsu(2016)]{Hsu2016role}
Raymond~K Hsu and Chi-yuan Hsu.
\newblock The role of acute kidney injury in chronic kidney disease.
\newblock \emph{Seminars in Nephrology}, 36\penalty0 (4):\penalty0 283--292,
  2016.
\newblock \doi{10.1016/j.semnephrol.2016.05.005}.

\bibitem[Montgomery(2009)]{Montgomery:2009}
Douglas~C. Montgomery.
\newblock \emph{Introduction to Statistical Quality Control, 6th Edition}.
\newblock John Wiley \& Sons, Jefferson City, MO, 2009.

\bibitem[Mortarino(2010)]{mortarino2010duncan}
Cinzia Mortarino.
\newblock Duncan's model for {\={x}}-control charts: Sensitivity analysis to
  input parameters.
\newblock \emph{Quality and Reliability Engineering International}, 26\penalty0
  (1):\penalty0 17--26, 2010.
\newblock \doi{10.1002/qre.1026}.

\bibitem[Zempl\'{e}ni et~al.(2004)Zempl\'{e}ni, V\'{e}ber, Duarte, and
  Saraiva]{Zempleni+etal:2004}
Andr\'{a}s Zempl\'{e}ni, Mikl\'{o}s V\'{e}ber, Belmiro Duarte, and Pedro
  Saraiva.
\newblock Control charts: a cost‐optimization approach for processes with
  random shifts.
\newblock \emph{Applied Stochastic Models in Business and Industry},
  20\penalty0 (3):\penalty0 185--200, 2004.
\newblock \doi{10.1002/asmb.521}.

\bibitem[Zhu and Park(2013)]{Zhu+Park:2013}
Weicheng Zhu and Changsoon Park.
\newblock \texttt{edcc}: An \texttt{R} package for the economic design of the
  control chart.
\newblock \emph{Journal of Statistical Software}, 52\penalty0 (9):\penalty0
  1--24, 2013.
\newblock \doi{10.18637/jss.v052.i09}.

\end{thebibliography}

\vspace{0.5cm}
	
	\noindent\textbf{Bal\'{a}zs Dobi and Andr\'{a}s Zempl\'{e}ni}\\
	Department of Probability Theory and Statistics, E\"{o}tv\"{o}s Lor\'{a}nd University\\
	Budapest,
		Hungary\\
	{\tt dobibalazs@inf.elte.hu}\\
	{\tt andras.zempleni@ttk.elte.hu}

\end{document}